\documentclass{amsart}
\usepackage{hyperref}
\usepackage{enumerate}
\newcommand*{\mailto}[1]{\href{mailto:#1}{\nolinkurl{#1}}}

\newtheorem{theorem}{Theorem}[section]
\newtheorem{lemma}[theorem]{Lemma}
\newtheorem{corollary}[theorem]{Corollary}

\numberwithin{equation}{section}

\newcommand{\R}{\mathbb{R}}
\newcommand{\Z}{\mathbb{Z}}

\newcommand{\floor}[1]{\lfloor#1 \rfloor}

\newcommand{\beq}{\begin{equation}}
\newcommand{\eeq}{\end{equation}}
\newcommand{\bal}{\begin{align}}
\newcommand{\eal}{\end{align}}
\newcommand{\nn}{\nonumber}

\newcommand{\si}{\sigma}
\newcommand{\pa}{\partial}

\newcommand{\la}{\lambda}

\newcommand{\ti}{\tilde}

\numberwithin{equation}{section}

\begin{document}

\title[On the mKdV Equation with Steplike Finite-Gap Initial Data]{On the Cauchy Problem for the
modified Korteweg--de Vries Equation with Steplike Finite-Gap Initial Data}

\author[I. Egorova]{Iryna Egorova}
\address{B. Verkin Institute for Low Temperature Physics\\
47 Lenin Avenue\\61103 Kharkiv\\Ukraine}
\email{\mailto{iraegorova@gmail.com}}

\author[G. Teschl]{Gerald Teschl}
\address{Faculty of Mathematics\\ University of Vienna\\
Nordbergstrasse 15\\ 1090 Wien\\ Austria\\ and\\ International
Erwin Schr\"odinger
Institute for Mathematical Physics\\ Boltzmanngasse 9\\ 1090 Wien\\ Austria}
\email{\mailto{Gerald.Teschl@univie.ac.at}}
\urladdr{\url{http://www.mat.univie.ac.at/~gerald/}}

\thanks{Research supported by the Austrian Science Fund (FWF) under Grant No.\ Y330.}
\thanks{Proceedings of the International Research Program on Nonlinear PDE, H. Holden and K. H. Karlsen (eds), 151--158, Contemp. Math. {\bf 526}, Amer. Math. Soc., Providence (2010)}

\keywords{mKdV, inverse scattering, finite-gap background, steplike}
\subjclass[2000]{Primary 35Q53, 37K15; Secondary 37K20, 81U40}

\begin{abstract}
We solve the Cauchy problem for the modified Korteweg--de Vries equation with
steplike quasi-periodic, finite-gap initial conditions under the assumption
that the perturbations have a given number of derivatives and moments finite.
\end{abstract}

\maketitle

\section{Introduction}

The purpose of the present paper is to investigate the Cauchy problem for the
modified Korteweg--de Vries (mKdV) equation
\beq\label{mKdV}
v_t(x,t) = -v_{xxx}(x,t) + 6 v(x,t)^2 v_x(x,t), \qquad v(x,0)=v(x),
\eeq
(where subscripts denote partial derivatives as usual) for the case of steplike
initial conditions $v(x)$. More precisely, we will assume
that $v(x)$ is asymptotically close to (in general) different real-valued, quasi-periodic, finite-gap potentials
$u_\pm(x)$ in the sense that
\beq\label{hypo}
\pm \int_0^{\pm \infty} \left|
\frac{d^n}{dx^n}\big( v(x) - u_\pm(x)\big) \right| (1+|x|^{m_0})dx <\infty,
\quad 0\leq n\leq n_0+1,
\eeq
for some positive integers $m_0, n_0$. Here by quasi-periodic, finite-gap potentials
we mean algebro-geometric, quasi-periodic, finite-gap potentials which arise naturally as
the stationary solutions of the mKdV hierarchy as discussed in \cite{GH}. If \eqref{hypo} holds for all $m_0, n_0$
we will call it a Schwartz-type perturbation.

If $u_\pm=0$ this problem is of course well understood, but for non-decaying initial conditions 
the only result we are aware of is the one by Kappeler, Perry, Shubin, and Topalov \cite{KPST}.
In order to solve the Cauchy problem for the mKdV equation \eqref{mKdV} with initial data satisfying
\eqref{hypo} for suitable $m_0,n_0$, our main ingredient will be the corresponding result for the
KdV equation \cite{ET}, \cite{EGT} combined with the Miura transform.

Next, let us state our main result.
Denote by $C^n(\R)$ the set of functions $x\in\R \mapsto q(x)\in \R$ which have $n$ continuous
derivatives with respect to $x$  and by $C^n_k(\R^2)$ the set of functions $(x,t)\in\R^2 \mapsto q(x,t)\in \R$
which have $n$ continuous derivatives with respect to $x$ and $k$ continuous derivatives with respect to $t$.

\begin{theorem}\label{theor1}
Let $u_\pm(x,t)$ be two real-valued, quasi-periodic, finite-gap solutions of the mKdV equation
corresponding to arbitrary quasi-periodic, finite-gap initial data $u_\pm(x)=u_\pm(x,0)$.
Let $m_0\geq 8$ and $n_0\geq m_0+5$ be fixed natural numbers.

Suppose, that $v(x)\in C^{n_0+1}(\R)$ is a real-valued function such that
\eqref{hypo} holds. Then there exists a unique classical solution $v(x,t)\in C^{n_0 - m_0-1}_1(\R^2)$
of the initial-value problem for the mKdV equation \eqref{mKdV} satisfying
\beq\label{decayv}
\pm \int_0^{\pm \infty} \left| \frac{\pa^n}{\pa x^n} \big( v(x,t) -
u_\pm(x,t)\big) \right| (1+|x|^{\floor{\frac{m_0}{2}} - 4})dx
<\infty, \quad n \leq n_0 - m_0 -1,
\eeq
for all $t\in\R$. Here $\floor{x}= \max \{ n\in\Z | n \leq x\}$ is the usual floor function.
\end{theorem}

In particular, this theorem shows that the mKdV equation
has a solution within the class of steplike Schwartz-type perturbations of finite-gap potentials:

\begin{corollary}
Let $u_\pm(x,t)$ be two real-valued, quasi-periodic, finite-gap solutions of the mKdV equation
corresponding to arbitrary quasi-periodic, finite-gap initial data $u_\pm(x)=u_\pm(x,0)$.
In addition, suppose, that $v(x)$ is a steplike Schwartz-type perturbations of $u_\pm(x)$.
Then the solution $v(x,t)$ of the initial-value problem for the mKdV equation \eqref{mKdV}
is a steplike Schwartz-type perturbations of $u_\pm(x,t)$ for all $t\in\R$.
\end{corollary}

For a unique continuation result within this class of solutions we refer to \cite{ET2}.

\section{The KdV equation with steplike finite-gap initial data}

As a preparation we recall some basic facts on the Cauchy problem for the
KdV equation
\beq\label{KdV}
q_t(x,t) = -q_{xxx}(x,t) + 6 q(x,t) q_x(x,t), \qquad q(x,0)= q(x),
\eeq
for the case of steplike initial conditions $q(x)$ from \cite{ET}, \cite{EGT}.
More precisely, we will assume that $q(x)$ is asymptotically close to (in general)
different quasi-periodic, finite-gap potentials $p_\pm(x)$ in the sense that
\beq\label{2.111}
\pm \int_0^{\pm \infty} \left|
\frac{d^n}{dx^n}\big( q(x) - p_\pm(x)\big) \right| (1+|x|^{m_0})dx <\infty,
\quad 0\leq n\leq n_0,
\eeq
for some positive integers $m_0, n_0$. The main result reads as follows

\begin{theorem}[\cite{ET}]\label{thmKdV}
Let $p_\pm(x,t)$ be two real-valued, quasi-periodic, finite-gap solutions of the KdV equation
corresponding to arbitrary quasi-periodic, finite-gap initial data $p_\pm(x)=p_\pm(x,0)$.
Let $m_0\geq 8$ and $n_0\geq m_0+5$ be fixed natural numbers.

Suppose that $q(x)\in C^{n_0}(\R)$ is a real-valued function such that
\eqref{2.111} holds. Then there exists a unique classical solution $q(x,t)\in C^{n_0 - m_0-2}_1(\R^2)$
of the initial-value problem for the KdV equation \eqref{KdV} satisfying
\beq \label{1.71}
\pm \int_0^{\pm \infty} \left| \frac{\pa^n}{\pa x^n} \big( q(x,t) -
p_\pm(x,t)\big) \right| (1+|x|^{\floor{\frac{m_0}{2}} - 2})dx
<\infty, \quad n \leq n_0 - m_0 -2,
\eeq
and
\beq \label{1.71t}
\pm \int_0^{\pm \infty} \left| \frac{\pa}{\pa t} \big( q(x,t) -
p_\pm(x,t)\big) \right| (1+|x|^{\floor{\frac{m_0}{2}} - 2})dx
<\infty,
\eeq
for all $t\in\R$.
\end{theorem}

In order to invert the Miura transform we will also need the solutions of the associated
Lax system.

Introduce the Lax operators corresponding to the finite-gap solutions $p_\pm(x,t)$,
\begin{align}\nn
L_\pm(t) &= -\pa_x^2 + p_\pm(x,t),\\\label{Lop}
P_\pm(t) &=  -4\pa_x^3 + 6p_\pm(x,t)\pa_x +3 \pa_x p_\pm(x,t).
\end{align}
Then the time dependent Baker--Akhiezer functions $\hat\psi_\pm(\la,x,t)$ are the unique solutions of the
Lax system (\cite{BBEIM}, \cite{GH})
\begin{align}\nn
L_\pm(t)\hat\psi_\pm &= \la\hat\psi_\pm,\\ \label{LPpm}
\frac{\pa\hat\psi_\pm}{\pa t} &= P_\pm(t)\hat\psi_\pm,
\end{align}
which satisfy $\hat\psi_\pm(\la,.,t)\in L^2(0,\pm\infty)$ and are normalized according to $\hat\psi_\pm(\la,0,0)=1$.
We will denote by $\breve\psi_\pm(\la,.,t)$ the other branch which satisfies $\breve\psi_\pm(\la,.,t)\in L^2(0,\mp\infty)$.

Similarly, for a solution $q(x,t)$ of the KdV equation as in Theorem~\ref{thmKdV} define the Lax operators
$L(t)$ and $P(t)$ as in \eqref{Lop} but with $q(x,t)$ in place of $p_\pm(x,t)$.

\begin{lemma}\label{lemKdV}
Let $q(x,t)$ be a solution of the KdV equation as in Theorem~\ref{thmKdV}. Then there exist
unique solutions of the Lax system
\begin{align}\nn
L(t) \hat\phi_\pm &= \la \hat\phi_\pm,\\ \label{LP}
\frac{\pa\hat\phi_\pm}{\pa t} &= P(t)\hat\phi_\pm,
\end{align}
which satisfy $\hat\phi_\pm(\la,.,t)\in L^2(0,\pm\infty)$ and are normalized according to
\beq
\hat\phi_\pm(\la,x,t) = \hat\psi_\pm(\la,x,t) (1 + o(1)) \qquad\text{as}\quad x \to\infty.
\eeq
Moreover, we have
\beq\label{hphipos}
\hat\phi_\pm(\la,x,t) > 0 \qquad \text{for} \quad \la \leq \inf \si(L(t)),
\eeq
where $\si(L(t))=\si(L(0))$ denotes the spectrum of the operator $L(t)$ in $L^2(\R)$.
\end{lemma}

\begin{proof}
The first part follows from \cite[Lemma 5.1]{EGT}. To see \eqref{hphipos} recall that
the Weyl solutions of $L(t) \phi = \la \phi$ have no zeros for $\la < \inf \si(L(t))$ and
thus $\hat\phi_\pm(\la,x,t) > 0$ for $\la < \inf \si(L(t))$
since the same is true for $\hat\psi_\pm(\la,x,t)$. Moreover, by continuity we obtain
$\hat\phi_\pm(\la,x,t) \geq 0$ for $\la \leq \inf \si(L(t))$ and since (nonzero) solutions
of a second order equation can only have first order zeros, we obtain \eqref{hphipos}.
\end{proof}

The solutions $\hat\phi_\pm(\la,x,t)$ can also be represented with the help of the transformation operators as
\beq\label{phipmKpm}
\hat\phi_\pm(\la,x,t) = \hat\psi_\pm(\la,x,t)\pm\int_x^{\pm\infty} K_\pm(x,y,t)
\hat\psi_\pm(\la,y,t) dy,
\eeq
where $K_\pm(x,y,t)$ are real-valued functions that satisfy
\beq
K_\pm(x,x,t)=\pm\frac{1}{2}\int_x^{\pm\infty} (q(y,t)-p_\pm(y,t))dy.
\eeq
Moreover, as a consequence of \cite[(A.15)]{BET}, the following
estimate is valid
\begin{align}\nn
\left|\frac{\pa^{n+l}}{\pa x^n\pa y^l} K_\pm(x,y,t)\right|\leq &
C_\pm(x,t)\Big(Q_\pm(x+y,t) \\ \label{estKpm}
& {} +\sum_{j=0}^{n+l-1} \left|\frac{\pa^j}{\pa x^j}\big(q(\frac{x+y}{2},t)
-p_\pm(\frac{x+y}{2},t)\big)\right|\Big),
\end{align}
for $\pm y>\pm x$, where $C_\pm(x,t)=C_{n,l,\pm}(x,t)$ are continuous positive
functions decaying as $x\to\pm\infty$ and
\beq
Q_\pm(x,t):=
\pm\int_{\frac{x}{2}}^{\pm\infty} \big|q(y,t) - p_\pm(y,t)\big| dy.
\eeq
Finally we recall, that for $\la\leq \inf \si(L(t))$ the equation $L(t) \phi = \la \phi$ has
two minimal positive (also known as principal or recessive) solutions which are
uniquely determined (up to a multiple) by the requirement
\[
\pm \int_0^{\pm\infty} \frac{dx}{\phi_\pm(\la,x)^2} = \infty.
\]
For $\la = \inf \si(L(t))$ the two minimal positive solutions could be linearly dependent and
the  $L(t)-\la$ is called critical in this case (and subcritical otherwise). And positive solution
can be written as a linear combination of the two minimal positive solutions and in the
critical case there is only one positive solution up to multiples. We refer to (e.g.) \cite{GZ}
for further details.

In particular, Lemma~\ref{lemKdV} implies that for $\la \leq \inf \si(L(t))$ the solutions $\hat\phi_\pm(\la,x,t)$
are the two minimal positive solutions of $L(t) \phi = \la \phi$ and thus any positive solution
of this equation is a multiple of
\beq
\hat\phi_\si(\la,x,t) = \frac{1+\si}{2} \hat\phi_+(\la,x,t) + \frac{1-\si}{2} \hat\phi_-(\la,x,t), \qquad \si\in[-1,1].
\eeq

Finally, we also recall the following uniqueness result.

\begin{theorem}[\cite{ET}]\label{thmKdVuniq}
Let $p_\pm(x,t)$ be two real-valued, quasi-periodic, finite-gap solutions of the KdV equation
corresponding to arbitrary quasi-periodic, finite-gap initial data $p_\pm(x)=p_\pm(x,0)$.
Suppose $q(x,t)$ is a solution of the KdV Cauchy problem satisfying
\beq\label{conduniq}
\pm \int_0^{\pm \infty} \left( |q(x,t) - p_\pm(x,t)| + \left| \frac{\pa}{\pa t} \big( q(x,t) -
p_\pm(x,t)\big) \right| \right) (1+x^2)dx <\infty,
\eeq
then $q(x,t)$ is unique within this class of solutions.
\end{theorem}

\section{The Miura transformation}

Our key ingredient will be the Miura transform \cite{Mi} and its inversion (see also \cite{G1}, \cite{gsi},
\cite{gs}, \cite{gss} and the references therein). Let $v(x,t)$ be a (classical) solution of the mKdV equation
\beq
v_t(x,t) = -v_{xxx}(x,t) + 6 v(x,t)^2 v_x(x,t).
\eeq
More precisely we will assume that
\beq
v_t, v_x, \dots,v_{xxxx}, \quad\text{and}\quad v_{xt}
\eeq
exist and are continuous.

Then
\beq\label{miurav}
q_j(x,t) = v(x,t)^2 + (-1)^j v_x(x,t), \qquad j=0,1,
\eeq
are classical solutions of the KdV equation. Moreover,
\beq\label{defphipm}
\phi_j(x,t)= \exp\left( (-1)^j \int_0^x v(y,t)
dy +(-1)^j \int_0^t (2 v(0,s)^3 - v_{xx}(0,s) ds\right)
\eeq
is a positive solution of
\begin{align}
-\frac{\pa^2}{\pa x^2}\phi_j(x,t) + q_j(x,t) \phi_j(x,t) &=0,\\
\frac{\pa}{\pa t}\phi_j(x,t) - \big( (-1)^j 2 q_j(x,t) v(x,t) -
q_{j,x}(x,t) \big) \phi_j(x,t) &= 0.
\end{align}
In other words, $\phi_j(x,t)$ solves the Lax system
\beq \label{Laxqpm}
L_j(t) \phi_j =0, \qquad \frac{\pa}{\pa t} \phi_j =
P_j(t)\phi_j,
\eeq
where the operators $L_j(t)$ and $P_j(t)$ are
defined as in \eqref{Lop} but with $q_j(x,t)$, $j=0,1$, in place of
$p_\pm(x,t)$. All claims are straightforward to check.

Conversely, let $q_j(x,t)$ be a solution of the KdV equation and let
$\phi_j(x,t)$ be a positive solution of \eqref{Laxqpm}, then one
sees after a quick calculation that
\beq
v(x,t) = (-1)^j \frac{\pa}{\pa x} \log \phi_j(x,t)
\eeq
is a solution of the mKdV equation.

\section{Finite-gap solutions of the mKdV equation}

In this section we want to briefly look at quasi-periodic, finite-gap solutions
of the mKdV equation and their relation to the quasi-periodic, finite-gap solutions
of the KdV equation (see also \cite{G3}, \cite{GH}).

Let $u_\pm(x,t)$ be quasi-periodic, finite-gap solutions of the mKdV
equation. Fix a number $j=0$ or $j=1$ for the Miura transformation.
Then
\beq \label{miurau}
p_{\pm,j}(x,t) = u_\pm(x,t)^2 + (-1)^j u_{\pm,x}(x,t)
\eeq
are quasi-periodic, finite-gap solutions of the
KdV equation.  Moreover, it is well-known (see, for example,
\cite{gsi}), that $\inf \si(L_{\pm,j}(t))\geq 0$, where
$L_{\pm,j}(t)$ is defined by \eqref{Lop}. Therefore, a positive
solution $\psi_{\pm,j}(x,t)$ defined as in \eqref{defphipm} with
$u_\pm$ instead of $v$, must be a convex combination of the two
branches of the Baker--Akhiezer function $\hat\psi_{\pm,j}(0,x,t)$
and $\breve\psi_{\pm,j}(0,x,t)$ corresponding to $p_{\pm,j}(x,t)$, that is,
\beq
\psi_{\pm,j}(x,t) = (1-\alpha_{\pm,j}(t)) \hat\psi_{\pm,j}(0,x,t) +
\alpha_{\pm,j}(t) \breve\psi_{\pm,j}(0,x,t).
\eeq
Moreover, either $0$ is the lowest band edge of  $\si(L_{\pm,j})$, in which case
$\hat\psi_{\pm,j}(0,x,t)=\breve\psi_{\pm,j}(0,x,t)$ and
$\alpha_{\pm,j}(t)$ drops out, or $0$ is below the spectrum
$\si(L_{\pm,j})$, in which case we must have $\alpha_{\pm,j}(t)=0$
or $\alpha_{\pm,j}(t)=1$ (since otherwise $0$ would be an eigenvalue
of operator, corresponding to the potential $u_\pm(x,t)^2 - (-1)^j
u_{\pm,x}(x,t)$).

Since the converse is also true, all quasi-periodic, finite-gap solutions of the mKdV equation
arise in this way from quasi-periodic, finite-gap solutions of the KdV equation.

Moreover, by virtue of Theorem~\ref{thmKdVuniq} we can already show the following
result which proves the uniqueness part of Theorem~\ref{theor1}.

\begin{theorem}
Let $u_\pm(x,t)$ be quasi-periodic, finite-gap solutions of the mKdV
equation and $v(x,t)$ a solution of the Cauchy problem for the mKdV
equation as above such that $q_0(x,t)$ (or $q_1(x,t)$) satisfies
\eqref{conduniq}. Then $v(x,t)$ is unique within this class.
\end{theorem}

\begin{proof}
Let $v(x,t)$ and $\ti{v}(x,t)$ be two solutions corresponding to the
same initial condition $v(x,0)=\ti{v}(x,0)=v(x)$. Then, by
uniqueness for KdV, $q_0(x,t) =\ti{v}(x,t)^2 + \ti{v}_x(x,t)$.
Moreover, $\phi_0(x,t)$ and $\ti{\phi}_0(x,t)$ defined by
\eqref{defphipm} both solves \eqref{LP} and coincide for $t=0$.
Hence they are equal by \cite[Lem.~2.4]{EGT} and so are $v(x,t)$ and
$\ti{v}(x,t)$.
\end{proof}

\section{Proof of the main theorem}

Let $u_\pm(x,t)$ be two quasi-periodic, finite-gap solutions of the
mKdV equation and suppose $v(x,t)$ is a (classical) solution of the
mKdV equation. Then
\beq
q_j(x,t) = v(x,t)^2 + (-1)^j\, v_x(x,t)
\eeq
is a classical solution of the KdV equation and $ p_{\pm,j}(x,t)$,
defined by \eqref{miurau} are quasi-periodic, finite-gap solutions
of the KdV equation. Choose numbers $j_\pm\in \{0,1\}$ for the
Miura transform such that (compare \eqref{defphipm})
\begin{align}\nn
\psi_\pm(x,t) &=\hat\psi_{\pm,j_\pm}(0,x,t)\\ \label{defjpm}
&= \exp\left((-1)^{j_\pm} \int_0^x u_\pm(y,t)
dy +(-1)^{j_\pm} \int_0^t (2 u_\pm(0,s)^3 - u_{\pm,xx}(0,s) ds\right)
\end{align}
and thus
\beq\label{choicesi}
\frac{\pa}{\pa x} \psi_\pm(x,t) = (-1)^{j_\pm} u_\pm(x,t) \psi_\pm(x,t),
\eeq
which is possible by the considerations from the last section.

\begin{lemma}\label{lemphip}
Let $u_+(x,t)$ and $v(x,t)$ be as introduced above such that
\beq
\int_0^\infty \big( |v(x,t) - u_+(x,t)| + |v_t(x,t) - u_{+,t}(x,t)| \big) dx < \infty.
\eeq
Then
\beq
\phi_+(x,t) :=  \psi_+(x,t) \exp\left((-1)^{j_++1} \int_x^\infty
(v(y,t)-u_+(y,t)) dy\right)
\eeq
is a minimal positive solutions of $(-\pa^2_x + q_{j_+}(x,t)) \phi =0$. 
Moreover,
\begin{align}\label{phit}
\frac{\pa}{\pa x} \phi_+(x,t) &= (-1)^{j_+} v(x,t) \phi_+(x,t),\\\label{phix}
\frac{\pa}{\pa t} \phi_+(x,t) &= \big( (-1)^{j_+} 2 q_{j_+}(x,t) v(x,t) -
q_{j_+,x}(x,t)\big) \phi_+(x,t).
\end{align}
\end{lemma}

\begin{proof}
First of all note that $\psi_+(x,t)=\hat \psi_{+,j_+}(0,x,t)$ is the
minimal positive solutions of $L_{+,j_+} \psi =0$ and by our choice
of $j_+$ we have \eqref{choicesi} from which \eqref{phit} is
immediate. Similarly, \eqref{phix} follows after a straightforward
computation.
\end{proof}

Now we are ready to prove our main theorem: We begin with the
initial condition $v(x)$ and define
\beq
q(x)= v(x)^2 + (-1)^{j_+} v_x(x).
\eeq
By our assumptions \eqref{hypo} we infer that $q(x)$
satisfies \eqref{2.111}. Hence, by Theorem~\ref{thmKdV} there is a
corresponding solution $q(x,t)$ of the KdV equation and by
Lemma~\ref{lemKdV} associated solution $\hat\phi_+(\la,x,t):=
\hat\phi_{+,j_+}(\la,x,t)$.

Recall \eqref{defjpm} and define $\phi_+(x)$ by
\beq \phi_+(x) :=
\psi_+(x,0) \exp\left( (-1)^{j_++1} \int_x^\infty (v(y)-u_+(y,0))
dy\right)
\eeq
which, by Lemma~\ref{lemphip} is a minimal positive
solution of $L(0)$. Moreover, since
\beq
\phi_+(x) = \psi_+(x,0) (1+ o(1)) \quad\text{as}\quad x\to\infty
\eeq
we conclude
\beq
\phi_+(x) = \hat\phi_{+,j_+}(0,x,0).
\eeq
Consequently
\beq
v(x,t) = (-1)^{j_+} \frac{\pa}{\pa x} \log \hat\phi_{+,j_+}(0,x,t)
\eeq
is a solution of the mKdV equation which satisfies the initial condition
\beq
v(x,0) =(-1)^{j_+} \frac{\pa}{\pa x} \log \hat\phi_{+, j_+}(0,x,0) =
(-1)^{j_+} \frac{\pa}{\pa x} \log \phi_+(x) = v(x)
\eeq
as required.

To see \eqref{decayv} set $\phi_+(x,t):=\hat\phi_{+,j_+}(0,x,t)$ and
observe that from \eqref{phipmKpm}
\beq\label{phipsi}
\frac{\phi_+(x,t)}{\psi_+(x,t)} =1 + \int_x^\infty
K_+(x,y,t)\frac{\psi_+(y,t)}{\psi_+(x,t)} dy,
\eeq
and thus
\[
1/2<\frac{\phi_+(x,t)}{\psi_+(x,t)}<2
\]
for $ x>  x_0(t)$. Moreover, differentiating \eqref{phipsi} we
obtain
\begin{align} \nn
v(x,t) - u_+(x,t) & = \frac{\pa}{\pa x} \log \frac{\phi_+(x,t)}
{\psi_+(x,t)}\\
& = \frac{\psi_+(x,t)}{\phi_+(x,t)} \Bigg( - K_+(x,x,t)
\\ \nn & \qquad + \int_x^\infty \big( K_{+,x}(x,y,t) - u_+(x,t)
K(x,y,t)\big) \frac{\psi_+(y,t)}{\psi_+(x,t)} dy \Bigg)
\end{align}
which implies
\beq
|v(x,t) - u_+(x,t)| \leq C_+(t) \left( Q_+(2x,t) + \int_x^\infty Q_+(x+y,t) dy \right).
\eeq
The higher derivatives then follow in a similar fashion using
\[
\frac{\pa}{\pa x} \big(v(x,t) - u_+(x,t) \big) = q(x,t) - p_+(x,t)
- \left(\frac{\phi_{+,x}(x,t)}{\phi_+(x,t)}\right)^2 +
\left(\frac{\psi_{+,x}(x,t)}{\psi_+(x,t)}\right)^2.
\]
This shows \eqref{decayv} for the plus sign. To see it for the minus
sign, repeat the argument with $j_-$.

\bigskip
\noindent{\bf Acknowledgments.} We are very grateful to F. Gesztesy
for helpful discussions. G.T. gratefully acknowledges the stimulating
atmosphere at the Centre for Advanced Study at the Norwegian Academy of Science and Letters in Oslo
during June 2009 where parts of this paper were written as part of  the international
research program on Nonlinear Partial Differential Equations.


\begin{thebibliography}{99}
\bibitem{BBEIM}
E. D. Belokolos, A. I. Bobenko, V. Z. Enolskii, A. R. Its, and V. B. Matveev,
{\em Algebro Geometric Approach to Nonlinear Integrable Equations},
Springer, Berlin, 1994.
\bibitem{BET} A. Boutet de Monvel, I. Egorova, and G. Teschl,
{\em Inverse scattering theory for one-dimensional Schr\"odinger
operators with steplike finite-gap potentials}, J. d'Analyse Math.
{\bf 106:1}, 271--316, (2008).
\bibitem{ET} I. Egorova and G. Teschl, {\em On the Cauchy
problem for the Korteweg--de Vries equation with steplike finite-gap initial data II.
Perturbations with Finite Moments}, J. d'Analyse Math. (to appear).
\bibitem{ET2} I. Egorova and G. Teschl, {\em A Paley-Wiener theorem for periodic scattering
with applications to the Korteweg-de Vries equation}, Zh. Mat. Fiz. Anal. Geom. {\bf 6:1}, 21--33 (2010).
\bibitem{EGT} I. Egorova, K. Grunert, and G. Teschl, {\em On the Cauchy
problem for the Korteweg--de Vries equation with steplike finite-gap initial data I.
Schwartz-type perturbations}, Nonlinearity {\bf 22}, 1431--1457 (2009).
\bibitem{G1} F. Gesztesy, {\em On the modified Korteweg--de Vries equation}, in
Differential Equations with Applications in Biology, Physics, and Engineering,
139--183, Marcel Dekker, New York, 1991.
\bibitem{G3} F. Gesztesy, {\em Quasi-periodic, finite-gap solutions of the modified Korteweg--de Vries},
in Ideas and Methods in Mathematical Analysis, Stochastics, and Applications, 428--471,
Cambridge UP, Cambridge, 1992.
\bibitem{GH} F. Gesztesy and H. Holden, {\em Soliton Equations and
their Algebro-Geometric Solutions. Volume {I}: $(1+1)$-Dimensional
Continuous Models}, Cambridge Studies in Advanced Mathematics,
Vol. {\bf 79}, Cambridge University Press, Cambridge, 2003.
\bibitem{gsi} F. Gesztesy and B. Simon, {\em Constructing solutions of the mKdV-equation},
J. Funct. Anal. {\bf 89:1}, 53--60 (1990). 
\bibitem{gs} F. Gesztesy and R. Svirsky, {\em (m)KdV-Solitons on the background of
quasi-periodic finite-gap solutions}, Memoirs Amer. Math. Soc. {\bf 118},
No. 563 (1995).
\bibitem{gss} F. Gesztesy, W. Schweiger, and B. Simon, {\em Commutation methods applied to the mKdV-equation},
 Trans. Amer. Math. Soc. {\bf 324:2}, 465--525 (1991).
\bibitem{GZ} F. Gesztesy and X. Zhao, {\em On critical and subcritical Sturm-Liouville operators},
J. Funct. Anal. {\bf 98:2}, 311--345 (1991).
\bibitem{KPST} T. Kappeler, P. Perry, M. Shubin and P. Topalov,
 {\it Solutions of mKdV in classes of functions unbounded at infinity},
J. Geom. Anal. {\bf 18}, 443--477 (2008).
\bibitem{Mi} R. M. Miura, {\em Korteweg--de Vries equation and generalizations. I.
a remarkable explicit nonlinear transformation}, J. Math. Phys. {\bf 9}, 1202--1204 (1968).
\end{thebibliography}
\end{document}